\documentclass[twocolumn,preprintnumbers,amsmath,amssymb,superscriptaddress]{revtex4}
\usepackage{epsf}
\usepackage{graphicx}
\usepackage{sidecap}
\usepackage{color}
\def \beq {\begin{equation}}
\def \eeq {\end{equation}}
\begin{document}
	
\title {Effect of dilute magnetism in a topological insulator}
\author{Firoza~Kabir}\affiliation {Department of Physics, University of Central Florida, Orlando, Florida 32816, USA.}
\author{M.~Mofazzel~Hosen}\affiliation {Department of Physics, University of Central Florida, Orlando, Florida 32816, USA.}
\author {Xiaxin Ding} \affiliation {Idaho National Laboratory, Idaho Falls, Idaho 83415, USA.}
\author{Christopher Lane} 
\affiliation {Theoretical Division, Los Alamos National Laboratory, Los Alamos, New Mexico 87545, USA}
\affiliation {Center for Integrated Nanotechnologies, Los Alamos National Laboratory, Los Alamos, New Mexico 87545, USA}
\author{Gyanendra~Dhakal}\affiliation {Department of Physics, University of Central Florida, Orlando, Florida 32816, USA.}
\author{Yangyang Liu}\affiliation {Department of Physics, University of Central Florida, Orlando, Florida 32816, USA.}
\author{Klauss~Dimitri}\affiliation {Department of Physics, University of Central Florida, Orlando, Florida 32816, USA.}
\author{Christopher Sims}\affiliation {Department of Physics, University of Central Florida, Orlando, Florida 32816, USA.}
\author{Sabin Regmi}\affiliation {Department of Physics, University of Central Florida, Orlando, Florida 32816, USA.}
\author{Luis Persaud}\affiliation {Department of Physics, University of Central Florida, Orlando, Florida 32816, USA.}
\author{Yong Liu}\affiliation {Crystal Growth Facility, Institute of Physics, {\'E}cole Polytechnique F{\'e}d{\'e}rale de Lausanne, CH-1015 Lausanne, Switzerland.}
\affiliation{Ames Laboratory, US Department of Energy, Ames, Iowa 50011, USA.}
\author{Arjun K. Pathak} \affiliation {Department of Physics, SUNY Buffalo State, Buffalo, New York 14222, USA. }
\author{Jian-Xin Zhu} \affiliation {Theoretical Division, Los Alamos National Laboratory, Los Alamos, New Mexico 87545, USA}
\affiliation {Center for Integrated Nanotechnologies, Los Alamos National Laboratory, Los Alamos, New Mexico 87545, USA}
\author{Krzysztof Gofryk} \affiliation {Idaho National Laboratory, Idaho Falls, Idaho 83415, USA.}
\author{Madhab~Neupane} \affiliation {Department of Physics, University of Central Florida, Orlando, Florida 32816, USA.}

\date{today}

\begin{abstract}
	
\noindent {Three-dimensional topological insulators (TIs) have emerged as a unique state of quantum matter and generated enormous interests in condensed matter physics. The surfaces of a three dimensional (3D) TI are composed of a massless Dirac cone, which is characterized by the Z$_2$ topological invariant. Introduction of magnetism on the surface of TI is essential to realize the quantum anomalous Hall effect (QAHE) and other novel magneto-electric phenomena. Here, by using a combination of first-principles calculations, magneto-transport, angle-resolved photoemission spectroscopy (ARPES), and time resolved ARPES (tr-ARPES), we study the electronic properties of Gadolinium (Gd) doped Sb$_2$Te$_3$. Our study shows that Gd doped  Sb$_2$Te$_3$ is a spin-orbit-induced bulk band-gap material, whose surface is characterized by a single topological surface state. We further demonstrate that introducing diluted 4$f$-electron magnetism into the Sb$_2$Te$_3$ topological insulator system by the Gd doping creates surface magnetism in this system. Our results provide a new platform to investigate the interaction between dilute magnetism and topology in doped topological materials.}
	
\end{abstract}
\date{\today}
\maketitle
\noindent
\noindent
\noindent In topological quantum materials (TQMs), the interplay of magnetism and topology can give rise to profoundly alluring phenomena including quantum anomalous Hall effect (QAHE), topological electromagnetic dynamics, and generate new states such as magnetic Dirac and Weyl semimetals, axion insulators etc. \cite{2L, 7L, 8L, 1L}. Although multiple intrinsic magnetic TQMs have been proposed theoretically \cite{ 12L, 13L}, their experimental realizations are rare, other than a few exceptions of magnetic Weyl, multifermionic magnetic material and nodal-line fermionic phase compounds \cite{MH, 19L, 20L, FK}. Recently, a new magnetic topological insulator has been theoretically proposed and later experimentally realized in thin-films and single crystals of MnBi$_2$Te$_4$ \cite{24L, 23L, 22L, 250L}. This system possesses a long-range antiferromagnetic (AFM) order and contributes a pragmatic platform to understand many compelling phenomena and phases, such as quantized anomalous Hall effect \cite{26L}, axion insulator phase \cite{Sabin, 27L}, high number Chern insulator \cite{28L}, and AFM TIs \cite{24L}. Despite these recent developments, there are still several mysteries and challenges concerning the topological electronic structure and its interplay with the magnetism. Specifically, there is little information on how topological properties are effected by dilute magnetism and how magnetic dopants can modify the electronic structure of a material. Time reversal symmetry in topological materials can be broken by doped magnetic impurities, yielding gap at the Dirac point \cite{cr4, cr3}. Earlier attempts have been made to investigate the doped magnetic materials or the proximity effect in magnetic heterostructure of topological materials and magnetic insulators, in which the magnetic response is weak \cite {Bo13, Bo16}. Typically, increased levels of dopants may raise the exchange field, but lower the sample quality and reduce electronic mobility \cite{H3, Bo27, Bo26, Bo28, Bo29}.  Consequently, QAHE has been realized at very low temperatures \cite{Bo13,Bo16}. These shortcomings have dragged behind the pace of the advancement of these materials for potential applications. Therefore, to generate more robust application platforms, we must find magnetic dopants that promote strong exchange fields at very low concentrations.   \\
\noindent To date, doped TIs have been studied by doping 3\textit{d} transition metals such as V, Cr, Mn, Fe, and Cu, where the doping generates various magnetic properties such as ferromagnetism, antiferromagnetism, and superconductivity \cite{cr8, cr10, cr7, cr9, cr11}. But to realize the QAHE, very low temperatures are needed to enhance the magnetic moments and suppress bulk dissipation channels, therefore improved materials are required \cite {Mo25}. To address this challenge, we utilize magnetism of 4\textit{f}-electron elements as an alternative to 3\textit{d} transition metals to dope the topological material. Due to their well-shielded 4\textit{f} shell, the high moment rare earth (RE) (especially gadolinium) ions are expected to behave like localized magnetic moments in the host matrix, leading to an overall paramagnetic behavior in the doped system \cite{Mo26}. However, measurements of RE-doped systems as well as pure RE metals have revealed a variety of magnetic properties, which are often complex and unpredictable \cite{Mo26}. Previous studies of doped 3D TIs have indicated that the size of the Dirac gap increases with the size of the magnetic moment and increased gap size provides greater flexibility for exploring magnetic TI physics \cite{Mo9}. Hence, large magnetic moment doping studies have already been conducted by utilizing Dy or Ho with moments 2-3 times larger than those typically observed in 3\textit{d} transition metal-doped systems \cite{Dy}.\\ 
\noindent Although some RE-doped systems such as Gd doped Bi$_2$Te$_3$ and Bi$_{1.09}$Gd$_{0.06}$Sb$_{0.85}$Te$_3$ have been studied by ARPES to confirm the existence of the topological surface state (TSS) \cite{Mo29, kimura}, to date, no study of Gd doped Sb$_2$Te$_3$ has been performed to probe the effect of dilute magnetism in it with its basic transport or electronic properties. Therefore, we utilize ARPES, time resolved ARPES (tr-ARPES), thermodynamic, and magneto-transport experiments, complemented by first principle calculations to investigate both the bulk and surface states of Gd$_{0.01}$Sb$_{1.99}$Te$_3$. The low concentration of Gd atoms in our compound, not only introduces dilute magnetism but also compensates n-type doping from vacancy and anti-site defects. The low-temperature electrical resistivity of Gd$_{0.01}$Sb$_{1.99}$Te$_3$ reveals a subtle local antiferromagnetic ordering below 2.4 K. For the first time, we report the direct observation of single Dirac cone at the $\Gamma$ point in the dilute magnetic topological material Gd$_{0.01}$Sb$_{1.99}$Te$_3$ by scanning over the entire Brillouin zone of this crystal using ARPES, tr-ARPES, and first-principles calculations. Our findings would initiate a new platform to study the interplay between dilute magnetism and topology in doped topological materials.\\  



\noindent\textbf{RESULTS}\\
\noindent{\textbf{Crystal structure and sample characterization}}\\
The Gd doped Sb$_2$Te$_3$ crystals were grown by a slow cooling technique (see methods for details). The trigonal crystal structure (R3m, $\#$166) of Sb$_2$Te$_3$ and 3D bulk Brillouin zone along with its projection onto the [001] surface with the high symmetry points (marked) are presented in fig. 1(a) and 1(b), respectively. In fig. 1 (c), the theoretically obtained bulk band structure of Gd doped Sb$_2$Te$_3$ is shown. A direct band gap of 0.18 eV is  observed at the $\Gamma$ high symmetry point of the Brillouin zone (BZ), with the remaining valence bands in the range of 0.45 - 0.75 eV below the Fermi level. Noticeably, at the $\Gamma$ point, the hole-like valence band is robust against the Gd-doping, whereas the well localized Gd-4\textit{f} states produce a flat Gd-4\textit{f} impurity level cuts through and hybridize with the electron-like conduction band, producing a narrow gap. The first-principles calculations were carried out on a 500 atom unit cell to capture the dilute Gd concentration and the AFM coupling between pairs of impurities (see supplementary figure 1 for more detail). Additionally, the system was treated using spin-orbit coupling and full non-collinear magnetism for all atomic species.

\noindent
\begin{figure*}	
	\includegraphics[width=18 cm]{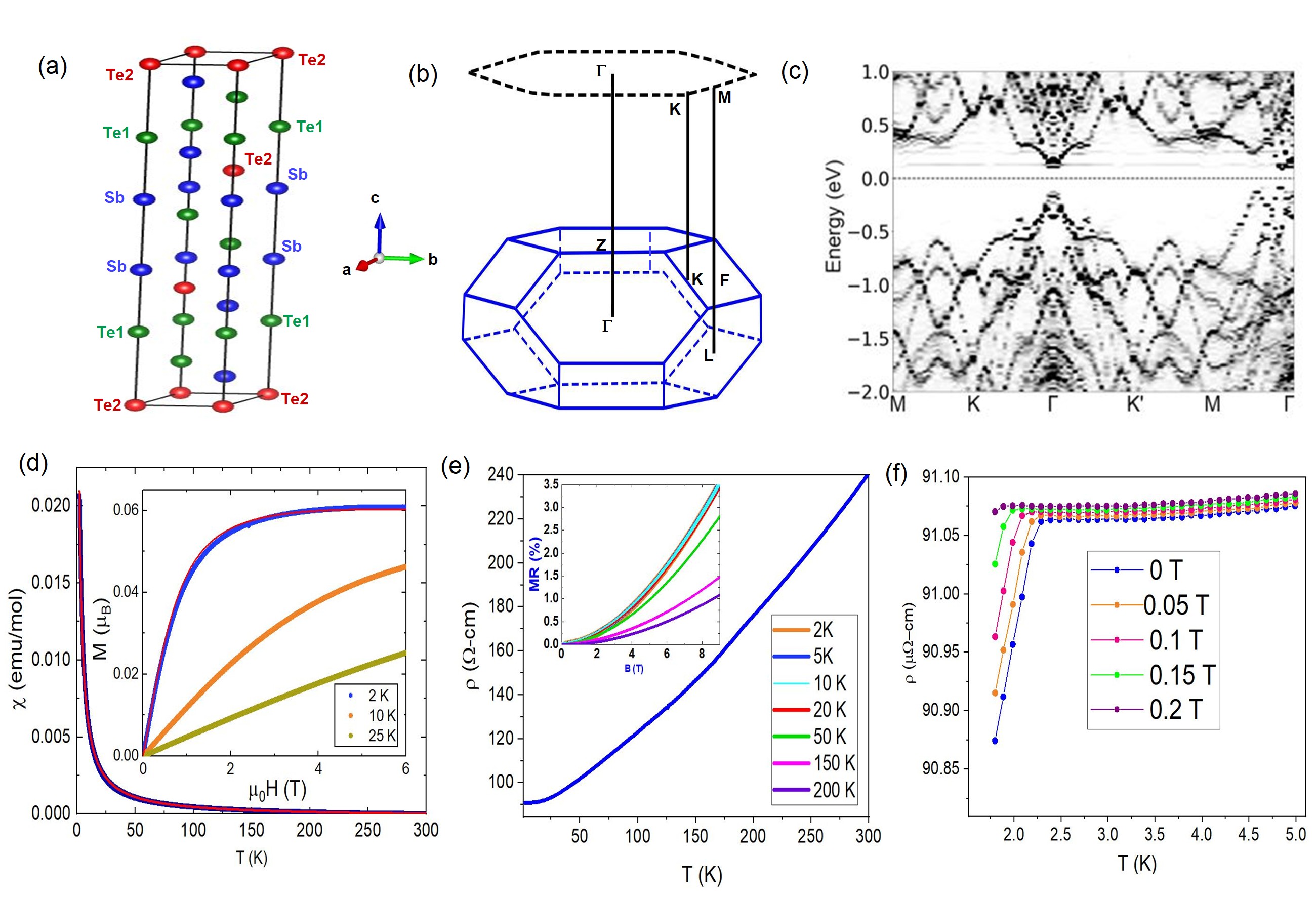}
	\caption{{Crystal structure and sample characterization. (a) Crystal structure of Sb$_2$Te$_3$. (b) Projection of the 3D bulk brillouin zone onto the [001] surface, high symmetry points are labeled. (c) First principles calculations of the bulk band of Gd$_{0.01}$Sb$_{1.99}$Te$_3$. The hole-like valence band is robust against the Gd-doping, whereas the Gd-4\textit{f} impurity level is hybridized with the electron-like conduction band, to produce a small gap. (d) Magnetic susceptibility as a function of temperatures together with a Curie-Weiss model shown by the solid red line (see the text). Inset shows the magnetization vs. magnetic field measured at different temperatures. The solid red line is a fitting curve.  (e) The electrical resistivity as a function of the temperature. Inset shows the magnetoresistivity taken at various magnetic fields. (f) The resistivity at low-temperatures measured at various magnetic fields.}}
\end{figure*}
\begin{figure*}
	\centering
	\includegraphics[width=18 cm]{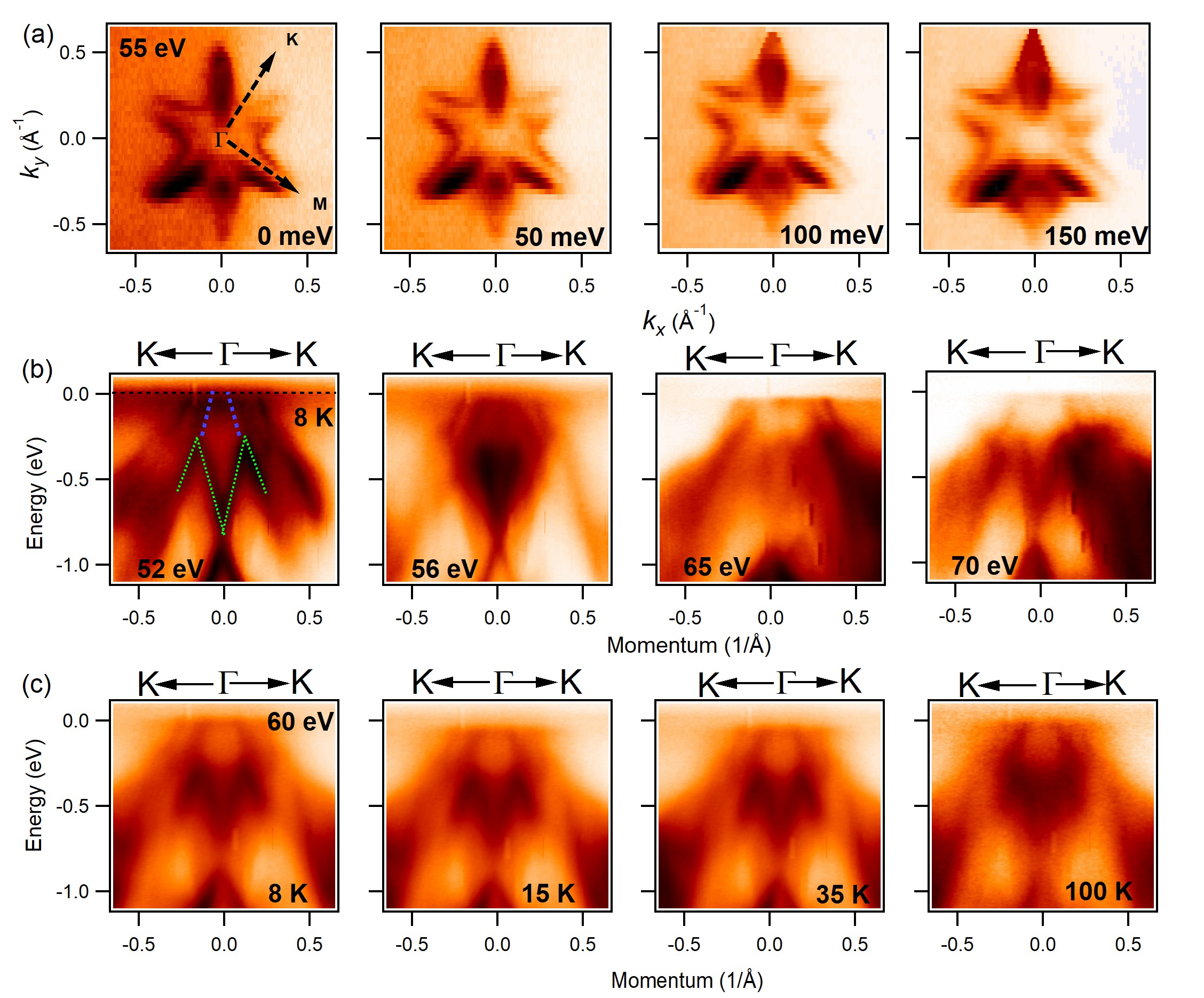}
	\caption{{Fermi surface, constant energy contours and  dispersion maps of Gd$_{0.01}$Sb$_{1.99}$Te$_3$}. (a) Fermi surface at photon energy 55 eV and corresponding constant energy contour plots. High symmetry points and binding energies are noted in the plots. (b) Photon energy dependent dispersion maps along the K-$\Gamma$-K direction. (c) Temperature dependent band dispersion of  Gd$_{0.01}$Sb$_{1.99}$Te$_3$ along the K-$\Gamma$-K direction. Photon energies and temperatures are noted on the plots.  All the experimental data were collected at ALS beamline 10.0.1.}
\end{figure*}
\begin{figure*}
	\centering
	\includegraphics[width=18 cm]{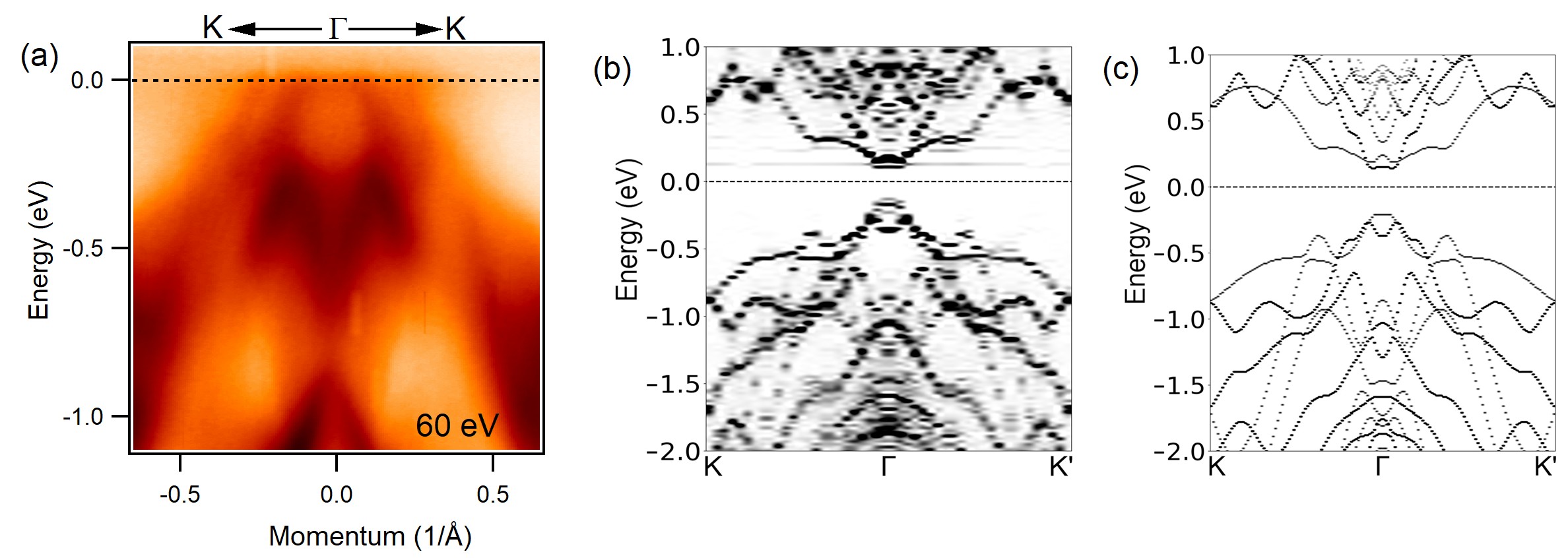}
	\caption{{Comparison of band structure of parent Sb$_2$Te$_3$ and Gd$_{0.01}$Sb$_{1.99}$Te$_3$.} (a) Experimental band dispersion map of Gd$_{0.01}$Sb$_{1.99}$Te$_3$ along the K-$\Gamma$-K direction. (b) DFT bulk band calculation of Gd$_{0.01}$Sb$_{1.99}$Te$_3$ along the K-$\Gamma$-K direction. (c) DFT bulk band calculation of parent Sb$_2$Te$_3$ along the K-$\Gamma$-K direction. All the experimental data were collected at ALS beamline 10.0.1.}
\end{figure*}

\noindent In order to determine the magnetic properties of Gd$_{0.01}$Sb$_{1.99}$Te$_3$, we have performed a detailed magnetization and magnetic susceptibility measurements. We have also used the magnetic measurements, together with energy dispersive spectroscopy (EDS) studies (see supplemental materials for more detail), to precisely determine the level of Gd content in our crystals. Fig. 1(d) shows the temperature dependence of the magnetic susceptibility of Gd$_{0.01}$Sb$_{1.99}$Te$_3$ single crystals with the applied magnetic field parallel to the c axis. Whilst the parent antimony telluride is itself a diamagnetic material \cite{TM}, our crystals display a characteristic Curie-Weiss magnetic susceptibility, confirming the presence of Gd in our Sb$_2$Te$_3$ crystals. The experimental data have been analyzed using a modified Curie-Weiss law (CW), $\chi(T)=C/(T-\theta)+\chi_0$, where C is the Curie constant, $\theta$ is the Curie-Weiss temperature, and $\chi_0$ is the temperature-independent contribution, most probably associated with the parent Sb$_2$Te$_3$. By fitting the CW law to the experimental data of our crystals yields, $\chi_0$ = -1.90 x 10$^{-4}$ emu/mol, $\theta$=~-1.19 K, and the Curie-Weiss parameter C = 0.0664 emu K mol$^{-1}$. The effective moment $\mu_{eff}$ = 0.7288 $\mu_{B}$/f.u. is calculated with $\chi_0$ taken out. The magnitude and negative sign of $\chi_0$ agrees well with the previously obtained magnetic susceptibility values of Sb$_2$Te$_3$, -2.3 x 10$^{-4}$ emu/mol \cite {NS}. Furthermore, the negative Curie temperature $\theta$=~-1.19 K suggests a possible tendency toward antiferromagnetic interactions among the Gd spins in this material. Assuming that the Curie term results only from localized trivalent Gd atoms, $\mu_{eff} = \sqrt{N\mu_{Gd}^2}$ ,where N and $\mu_{Gd}$ is the concentration of Gd atoms and the effective magnetic moment of the Gd atoms, respectively. Since the effective magnetic moment on the Gd$^{3+}$, $^8$S$_{7/2}$ states with theoretical J = 7/2, is $\mu_{Gd}$= 7.94 $\mu_{B}$, the concentration of Gd atoms is estimated to be 0.84\%. The inset of fig. 1(d) shows the field dependence of the magnetization of our sample measured at various temperatures. The isotherms show a characteristic Brillouin-like curvature expected for a Curie-Weiss paramagnet with saturation at high magnetic fields. The solid red curve is a fitting of the modified Brillouin function $M = Ng\mu_{B}JB_{J}(x) + \chi_0B$ at 2 K, where $B_J(x)$ is the Brillouin function with $x = \frac{g\mu_{B}JB}{k_BT}$. Here, N is estimated to be 0.89\%, which is in good agreement with the Gd concentration derived from the magnetic susceptibility data. The value $\sim$0.9\% is also very close to $\sim1\%$ obtained by the energy dispersive spectroscopy and scanning electron microscopy methods (see supplementary figure 6 for more detail) \cite{SF}. Therefore, we conclude that the concentration of Gd impurities in our sample to be 1\%, and we would use this value for the remainder of our paper.\\
The temperature dependence of the electrical resistivity of Gd$_{0.01}$Sb$_{1.99}$Te$_3$ shows a typical metallic behavior in the whole measured temperature range (fig. 1(e)). Inset of fig. 1(e) presents the magnetoresistance of the sample, at various temperatures. As seen from the figure, the magnetoresistivity is relatively small (about 3.5$\%$) and exhibits $\sim H^2$ dependence characteristic of metallic systems. Interestingly, at low temperatures, (fig. 1(f)), there is a weak anomaly in the $\rho(T)$ curve at around 2.4 K. This transition is shifted by applying a magnetic field as strong as 2 T. Though the negative value of $\theta_P$ might suggest the presence of long-range antiferromagnetic ordering in Gd$_{0.01}$Sb$_{1.99}$Te$_3$, a lack of clear anomalies in both $\chi(T)$ and $C_p(T)$ (see supplemental figure 2) indicate that the magnetic ordering is not a bulk phenomena. In general, a topological insulator doped with magnetic impurities can exhibit a long-range magnetic order on the surface, and such ordering can be independent of a bulk magnetic ordering \cite{Mo9, cr10}. Such ordering (with or without bulk magnetic order) can also lead to the breaking of TRS, that can result in a gap opening at the Dirac point making the surface state massive. Therefore, more studies are required to draw any firm conclusions on the nature of the low-temperature behavior in this material. 
\\
\noindent 
\textbf{Electronic structure}\\
\noindent To elucidate the electronic structure of Gd$_{0.01}$Sb$_{1.99}$Te$_3$, we measure the electronic dispersion maps and the Fermi surface of our sample using high-resolution ARPES. Fig. 2(a) reveals the Fermi surface map of Gd$_{0.01}$Sb$_{1.99}$Te$_3$ in the paramagnetic state (T = 8 K), with the constant energy contours at various binding energies with photon energy of 55 eV. A star shaped Fermi pocket around the center of the BZ ($\Gamma$ point) is observed and by moving towards the higher binding energy, we find that the star shape gradually evolves into a flower shaped pocket. The density of states at \textit{E}$_F$ is distributed over this star shaped pocket enclosing the $\Gamma$ point. A circular hole like feature is also observed at the $\Gamma$ point, which is the position of the topological surface state (evolving with the increasing binding energy). Additionally, the enlarging shape of the central pocket as binding energy increases confirms the hole like nature of the band around the $\Gamma$ point. 

\noindent In order to reveal the nature of the bands along various high symmetry directions, we present photon energy and temperature dependent ARPES-measured dispersion maps in fig. 2(b) and 2(c), where the photon energies are noted in the plots. Fig. 2(b) shows the measured dispersion maps along the K-$\Gamma$-K direction at various photon energies. From the leftmost figure of 2(b), we observe one hole like band in the vicinity of the Fermi level (blue dotted line) and one M-shaped band (green dotted line) around 250 meV below the Fermi level. A Dirac cone like feature is revealed around 750 meV below the chemical potential. Moving from left to right panels in figure 2(b), the relative intensity of various features changes with photon energy due to the matrix element effect, but the positions of the hole-like band, the M shaped band, and the Dirac cone like feature at 750 meV binding energy do not evolve and look exactly the same.  
From our overall observation, we can conclude that the linearly dispersive states close to the Fermi level originate from the surface. To further clarify the structure and the hole like band near the Fermi level, the second derivative plot is presented in supplementary figure 3 (a) \cite{SF}, which more precisely confirms the presence of one hole like band close to the Fermi level along K-$\Gamma$-K direction. Additionally, dispersion maps along the M-$\Gamma$-M direction with its second derivative plots are presented in supplementary figure 3(b), where one linear hole like band is observed in the vicinity of the Fermi level. From these linear dispersion maps presented in fig. 2(b), we observe a single Dirac cone in the vicinity of the Fermi level. In the further section, we will describe the process that we have used to observe the single Dirac cone using our static ARPES and tr-ARPES data. 

\noindent To test the robustness of the observed surface states of Gd$_{0.01}$Sb$_{1.99}$Te$_3$, we carry out a series of temperature-dependent ARPES measurements at photon energy of 60 eV. Fig. 2(c) shows the temperature dependent dispersion maps of Gd$_{0.01}$Sb$_{1.99}$Te$_3$ along the K-$\Gamma$-K direction of the Brillouin zone. We notice that in the paramagnetic state, Gd$_{0.01}$Sb$_{1.99}$Te$_3$ features very similar energy-momentum dispersions at various higher temperatures ranging from 8K to 100K, establishing the robustness of the observed surface states in this magnetic system. \\

\noindent\textbf{Observation of single Dirac cone}\\
 From the electronic structure measurement of Gd$_{0.01}$Sb$_{1.99}$Te$_3$, we observe the surface state in two ways, one way is through static ARPES and the other is noticed from the time-resolved ARPES (tr-ARPES). To find the position of the Dirac point above the Fermi level, first we focus on the dispersion map of Gd$_{0.01}$Sb$_{1.99}$Te$_3$ in fig. 3(a), where we observe one hole-like band crossing the Fermi level. By extrapolation, the Dirac point is found to be located approximately around 160 meV above the Fermi level, which is comparable with the previously observed Dirac point position for primitive Sb$_2$Te$_3$ (180 meV) \cite{DP}. Hence, adding small amount of Gd made the position of the Dirac point closer to the Fermi level. Now, from the theoretically obtained bulk band electronic structure for both doped and pristine Sb$_2$Te$_3$ compounds, (fig. 3(b) and 3(c), respectively) one can see that the electron-like bands of the Gd doped system are shifted closer to the Fermi level with respect to those of the pristine system (to compare with the surface state of the pristine system, see supplementary figure 1(c)) \cite{SF}. With the hole-like bands aligned for both doped and pristine systems, we are able to identify that the band gap is narrower in the Gd doped Sb$_2$Te$_3$ as compared to pristine Sb$_2$Te$_3$.\\
 \begin{figure*}
 \centering
 \includegraphics[width=18 cm]{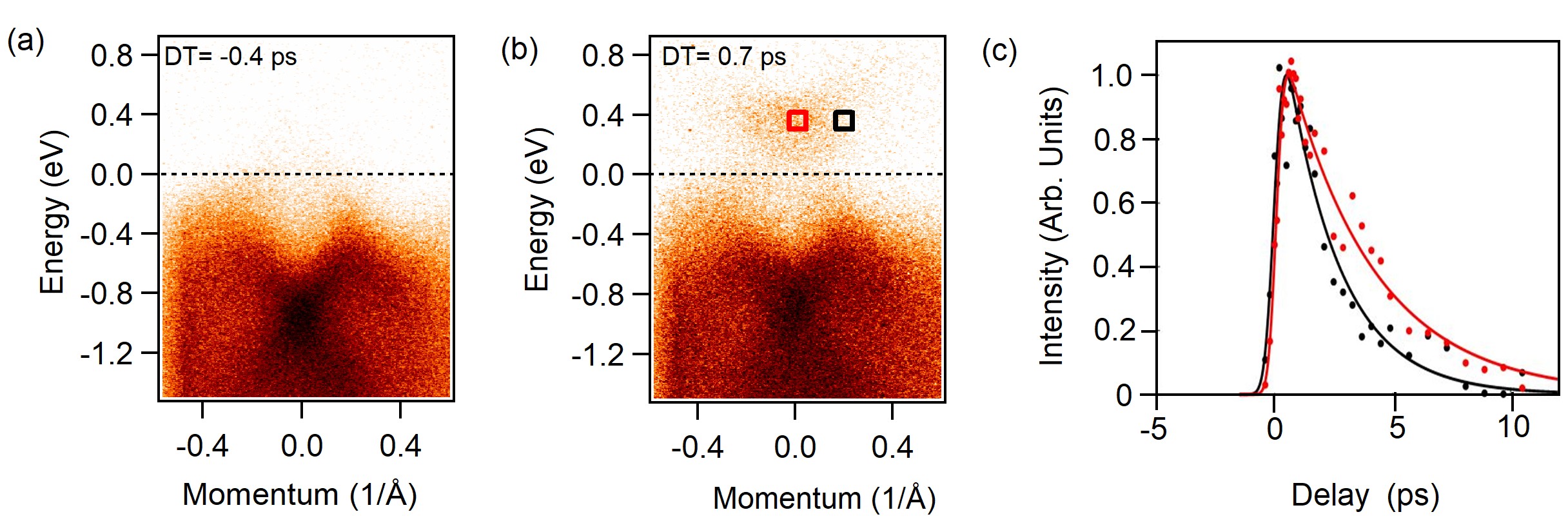}
 \caption{{tr-ARPES data of Gd$_{0.01}$Sb$_{1.99}$Te$_3$}. (a) Band dispersion of Gd$_{0.01}$Sb$_{1.99}$Te$_3$ at delay time (DT)= -0.4 ps. (b) tr-ARPES band dispersion of Gd$_{0.01}$Sb$_{1.99}$Te$_3$ at DT=0.7 ps. The black and red rectangles represent the integration window of transient photoemission intensity for the surface and the bulk states respectively, in Gd$_{0.01}$Sb$_{1.99}$Te$_3$. (c) Ultrafast evolution of the population of surface (black curve) and bulk state (red curve) in Gd$_{0.01}$Sb$_{1.99}$Te$_3$.}
 \end{figure*} 
 
 
 \noindent From these bulk band structure calculations, it is clear that a small amount of added Gd in Sb$_2$Te$_3$ (fig. 3(b)) did not make a huge change of the bulk bands compared to the pristine Sb$_2$Te$_3$ (fig 3(c)), the only difference is including Gd, the electron-like band at $\Gamma$ comes closer to the Fermi level. Again, from the surface state calculation of primitive Sb$_2$Te$_3$ (see Supplementary figure 1(c)) \cite{SF}, we already observe non trivial Dirac-like band at the $\Gamma$ high symmetry point. Hence, we can predict that, from Gd doped Sb$_2$Te$_3$ surface state calculation, we would be able to see a similar surface state close to the Fermi level, like that observed in the primitive Sb$_2$Te$_3$ system. Additionally, from the bulk band calculation of fig. 3(b), we find adding small amount of Gd do not disturb any major or minor bands compared to the bands of the pristine Sb$_2$Te$_3$ (fig 3(c)). Hence we can conclude that including Gd we would see the similar Dirac-like non- trivial surface state of Sb$_2$Te$_3$ at the $\Gamma$ point, where the electron like band would be closer to the Fermi level. \\~\\
 
\noindent \textbf{Time resolved ARPES measurement}\\
To reveal the existence of Dirac cone above Fermi level, we must go beyond the standard static ARPES techniques. Hence, we perform an experiment using  time- resolved ARPES technique. Unlike the ARPES technique, this technique can directly measure the non-equilibrium band structure of materials resulting from ultrafast excitation, thereby opening a new pathway to measure electron dynamics on femtosecond to picosecond timescales. Here, we use tr-ARPES with the pump-probe method to measure the unoccupied electronic states above \textit{E}$_F$. Fig. 4(a) shows dispersion map of Gd$_{0.01}$Sb$_{1.99}$Te$_3$ with delay time (DT) -0.4 ps, while fig. 4(b) presents the band structure with pump and probe pulses overlapped, with delay time (DT) 0.7 ps. Additionally, tr-ARPES spectra  at representative delay time (DT) values are shown in supplementary figure 5. After careful observation, from fig. 4(b), one can see Dirac-like feature above the Fermi level (at around 155 meV), which directly confirms our predicted Dirac point position from ARPES measurement. The black and red rectangles represent the integration window of transient photoemission intensity of surface and  bulk states in Gd$_{0.01}$Sb$_{1.99}$Te$_3$ respectively. Fig. 4(c) presents the integrated intensity measured as a function of the pump-probe delays for the bulk and surface state of the crystal. To analyze the decay lifetime of the excited state, we again fit the integrated signal to an exponentially-modified Gaussian distribution. The decay timescale of 2.29 ps for the surface state and 3.4 ps for the bulk are compatible with previous measurements of relaxation dynamics of topological insulators using laser-based tr-ARPES and give us additional confidence in the presence of the Dirac state above the Fermi level in this material \cite{tr6, tr7, Neupane}.\\~\\

\noindent \textbf{DISCUSSION}\\ 
\noindent
We would like to discuss the possible gap opening at the Dirac point. Our data shows a small band gap between the conduction and valence bands (fig. 4(b)). By breaking time-reversal symmetry at the Dirac point of topological insulators, magnetic impurities could play an important role to open a band gap, which has been studied theoretically and experimentally \cite {Rader4, Rader11}. However, some previous studies show that various dopants and their concentration levels might also effect the  surface band gap, which could be associated with the impurity-dependent strength U, regardless of whether the dopant is magnetic or not \cite {Rader42, Rader43, kimura}. In our present study, we can see a gap in between the valence and conduction bands as spectral weight may diminish near the Dirac point \cite {H3, Rader}, however, Gd doping does not effect the bulk band inversion and the system remains topologically nontrivial. \\
In general, a topological insulator doped with magnetic impurities can exhibit a long-range magnetic order on the surface \cite{cr10,Mo9} and such ordering can also break the TRS. The low-temperature anomaly observed in the electrical resistivity data and the negative value of $\theta$, (paramagnetic Curie temperature from magnetic susceptibility data, fig 1(d)) might suggest the presence of antiferromagnetic ordering in Gd$_{0.01}$Sb$_{1.99}$Te$_3$, but lacking of clear anomalies in both $\chi(T)$ and $C_p(T)$ at low temperatures (see Supplementary figure 2) indicating that the magnetic ordering is not a bulk phenomena. Consequently a gap might open at the Dirac point on the surface of our system. However, more studies are needed to draw any firm conclusions on the nature of low-temperature behavior in this material.

\noindent In summary, we use detailed transport, electrical transport, ARPES, and tr-ARPES measurements together with DFT calculations to study the electronic structure of the Gd-doped Sb$_{2}$Te$_3$ topological insulator. The magnetic and transport measurements show that Gd doping gives rise to local magnetic moments with Curie-Weiss-like behavior of the magnetic susceptibility and show some indications of surface magnetic ordering at 2.4 K. We observe the surface state from the electronic structure measurement of Gd$_{0.01}$Sb$_{1.99}$Te$_3$ in two ways, one through static ARPES and the other is through tr-ARPES technique. The surface states are directly associated with the 4\textit{f}-electron magnetism of gadolinium. The approximate position of the Dirac point around 160 meV above the Fermi level is also confirmed by the tr-ARPES data. We show that adding a small amount of gadolinium introduces dilute magnetism into it, and that does not prevent the formation of the Dirac state in a relatively wide temperature range. Hence, our overall electronic structure reveals topological non-trivial surface state above the Fermi level in this material, which could be a promising platform to study the effect of dilute magnetism on the electronic structures and topology of similar doped materials. \\~\\

\noindent\textbf{METHODS}\\
\textbf{Crystals growth and characterizations}\\
Single crystals of Gd$_{0.01}$Sb$_{1.99}$Te$_3$  were grown by melting the stoichiometric mixture of elemental Gd, Sb and Te at temperature of 1123 K for 24 hours in a sealed vacuum quartz tube. The sample was then cooled down over a period of 48 hours until it reaches 893 K, and was stored at this temperature for additional 96 hours before quenching in liquid nitrogen. In this way, crystals of centimeter size with a shiny surface could efficiently be
attained. Magnetization, electrical resistivity (four-probe method), and heat capacity measurements have been performed using a Quantum Design DynaCool-9 system equipped with a 9 T superconducting magnet and VSM, ETO, and HCP options. The phase purity of the sample was confirmed, with no observation of contaminated phases, by a powder x-ray diffraction (XRD) method (PANalytical X’Pert PRO diffractometer). The XRD pattern was fitted well with the hexagonal
structure of Gd$_{0.01}$Sb$_{1.99}$Te$_3$ with space group R3m. 
\\~\\
\textbf {Electronic structure measurements}\\
Synchrotron-based ARPES measurements of the electronic structure of Gd$_{0.01}$Sb$_{1.99}$Te$_3$ were performed at ALS BL 10.0.1 with a Scienta R4000 hemispherical electron analyzer. The samples were cleaved in situ in an ultra high vacuum conditions (5x10\textsuperscript{-11} Torr) at 8 K. The energy resolution was set to be better than 20 meV and the angular resolution was set to be better than  0.2$^{\circ}$ for the synchrotron measurements.
The Gd$_{0.01}$Sb$_{1.99}$Te$_3$ specimens were found to be very durable and did not exhibit any signs of deterioration for the typical measurement period of 20 hours. A small sample piece cut from the crystal was mounted on a copper post. We then used silver epoxy to attach a ceramic post on the top of the sample. The whole set was then loaded into the measurement chamber for in  situ cleavage before the measurement.
\\~\\
\textbf{Time-resolved measurement}\\
The tr-ARPES setup at Laboratory For Advanced Spectroscopic Characterization of Quantum Materials (LASCQM), University of Central Florida consists of a moderately high power (20 W), high repetition rate (50 to 200 kHz) Yb:KGW laser and a hemispherical analyzer \cite{Yang}. The pump and probe photon energies are 1.2 and 21.8 meV, respectively. During the measurement, the time and energy resolutions were set to be 320 fs and \textless 50 meV, respectively. The delay stage (Newport, DL125), with a scan range of 0.8 ns and minimum step size of 0.5 fs, was used to control the time delay between the pump and probe pulses. The samples were cooled down to 77 K and the pump fluence was set to 2.7 mJcm$^{-2}$.\\

\noindent\textbf{Electronic structure calculations}\\
First-principles calculations were carried out using the pseudopotential projected augmented wave method \cite{THEORY1} implemented in the Vienna {\em ab initio} simulation package \cite{THEORY3, THEORY2} with an energy cutoff of 420 eV for the plane-wave basis set. Exchange-correlation effects were treated using the generalized gradient approximation (GGA) \cite{THEORY4}, where only the $\Gamma$-point was used to sample the Brillouin zone of the super cell crystal structure. A 4 $\times$ 4 $\times$ 1 super cell of Sb$_2$Te$_3$ was considered to achieve the dilute doping (2\%) of antiferromagnetically ordered Gd impurities. Experimental lattice parameters and atomic positions were used.  A total energy tolerance of 10$^{-6}$ eV was used in self-consisting the charge density. Spin orbit coupling effects were included in a self-consistent manner. To compare our super cell band dispersion to the ARPES spectra, we unfolded the bands into the primitive cell using BandUp \cite{THEORY5,THEORY6}. Further benchmarks were performed with the full-potential linearized augmented plane wave (FP-LAPW) method implemented in a WIEN2k package \cite{THEORY7}\\


\noindent
\\
\textbf{ACKNOWLEDGEMENTS}\\
M.N. is supported by the Air Force Office of Scientific Research under Award No. FA9550-17-1-0415 and the Center for Thermal Energy Transport under Irradiation, an Energy Frontier Research Center funded by the U.S. DOE, Office of Basic Energy Sciences. K.G. and X.D. acknowledge support from INL's LDRD program (19P45-019FP) and the DOE's Early Career Program. This work at Los Alamos was carried out under the auspices of the U.S. Department of Energy (DOE) National Nuclear Security Administration under Contract No. 89233218CNA000001 (C. L. and J.-X.Z.). The work was supported, in part, by the Center for Integrated Nanotechnologies, a DOE BES user facility, in partnership with the LANL Institutional Computing Program for computational resources. The work at Ames Laboratory was supported by the U.S. DOE, Office of Basic Energy Sciences under Contract No. DE-AC02-07CH11358 with Iowa State University. We thank Sung-Kwan Mo for beamline assistance at the LBNL.\\


\noindent \*\textbf{Correspondence} and requests for materials should be addressed to M.N. (Email: Madhab.Neupane@ucf.edu).

\end{document}